\documentclass{article}

\usepackage[preprint,nonatbib]{neurips_2024}

\usepackage[utf8]{inputenc} 
\usepackage[T1]{fontenc}    
\usepackage{hyperref}       
\usepackage{url}            
\usepackage{booktabs}       
\usepackage{amsfonts}       
\usepackage{nicefrac}       
\usepackage{microtype}      
\usepackage{xcolor}         
\usepackage{amsmath}
\usepackage{siunitx}
\usepackage[numbers,sort&compress]{natbib}
\usepackage[pdftex]{graphicx} 
\usepackage{algorithm}
\usepackage{algpseudocode}
\usepackage[symbol]{footmisc}
\usepackage{wrapfig}

\title{Optical Diffusion Models for Image Generation}

\author{
Ilker Oguz$^1$ \quad Niyazi Ulas Dinc$^1$ \quad Mustafa Yildirim$^1$ \quad Junjie Ke$^2$ \quad Innfarn Yoo$^2$ \\
\textbf{Qifei Wang}$^3$ \quad \textbf{Feng Yang}$^2$\footnotemark[1] \quad \textbf{Christophe Moser}$^1$\footnotemark[1] \quad \textbf{Demetri Psaltis}$^1$\footnotemark[1] \\
$^1$ École Polytechnique Fédérale de Lausanne  \quad $^2$ Google Research  \quad $^3$ Google\\
\texttt{\{ilker.oguz,niyazi.dinc,mustafa.yildirim,christophe.moser,demetri.psaltis\}@epfl.ch}\\
\texttt{\{junjiek,innfarn,qfwang,fengyang\}@google.com}\\
}

\begin{document}

\maketitle

\footnotetext[1]{Equal advising.}

\begin{abstract}
Diffusion models generate new samples by progressively decreasing the noise from the initially provided random distribution. This inference procedure generally utilizes a trained neural network numerous times to obtain the final output, creating significant latency and energy consumption on digital electronic hardware such as GPUs. In this study, we demonstrate that the propagation of a light beam through a semi-transparent medium can be programmed to implement a denoising diffusion model on image samples. This framework projects noisy image patterns through passive diffractive optical layers, which collectively only transmit the predicted noise term in the image. The optical transparent layers, which are trained with an online training approach, backpropagating the error to the analytical model of the system, are passive and kept the same across different steps of denoising.  Hence this method enables high-speed image generation with minimal power consumption, benefiting from the bandwidth and energy efficiency of optical information processing.

\end{abstract}

\section{Introduction}
Diffusion models create new samples that resemble their training sets by gradually undoing the diffusion process, which requires the learned reverse process to be applied numerous times \cite{sohl-dickstein_deep_2015}. While this method demonstrated unprecedented capabilities by producing highly realistic samples \cite{ho_denoising_2020,song_score-based_2020,nichol_improved_2021, jalal_robust_2021,rombach_high-resolution_2022,saharia_photorealistic_2022,saharia_image_2023}, it is also highly time-consuming and expensive in terms of energy consumption and computing resources since a large number of steps are required for generating each sample \cite{ho_denoising_2020}. This prolonged processing time not only limits accessibility but also contributes to a significant environmental footprint.

Currently, generating new samples with diffusion models relies on electronic, general-purpose computing hardware such as GPUs or TPUs. However, due to the repetitive nature of the reversal process required in this task, deploying specialized hardware instead of general-purpose ones could significantly enhance the efficiency of sampling. For instance, the use of ASICs in cryptocurrency mining for hashing algorithms has demonstrated substantial improvements in computational speed and energy efficiency \cite{kufeoglu_bitcoin_2019}. However, both GPUs and ASICs, among other electronic digital computers,  face the same challenges like heat dissipation, energy consumption, and the diminishing returns of Moore's Law, as transistors shrink and encounter quantum effects and physical limits that hinder further gains \cite{leiserson_theres_2020}. Therefore, exploring alternative computing modalities, such as optical computing—which offers high bandwidth and low loss—is increasingly important \cite{wetzstein_inference_2020}. Optical computing has already shown promise in various applications, including high-speed data transmission and real-time signal processing. Optical computing addresses the inefficiencies of electronic hardware by leveraging the inherent parallelism that light allows for the simultaneous processing of multiple data channels, significantly speeding up the computational process. Several optical neural networks have been reported to perform complex calculations at reduced latency and energy consumption compared to traditional electronic systems \cite{mcmahon_physics_2023-1}.  

\begin{figure}[h]
  \centering
  \includegraphics[width = 1.0\textwidth]{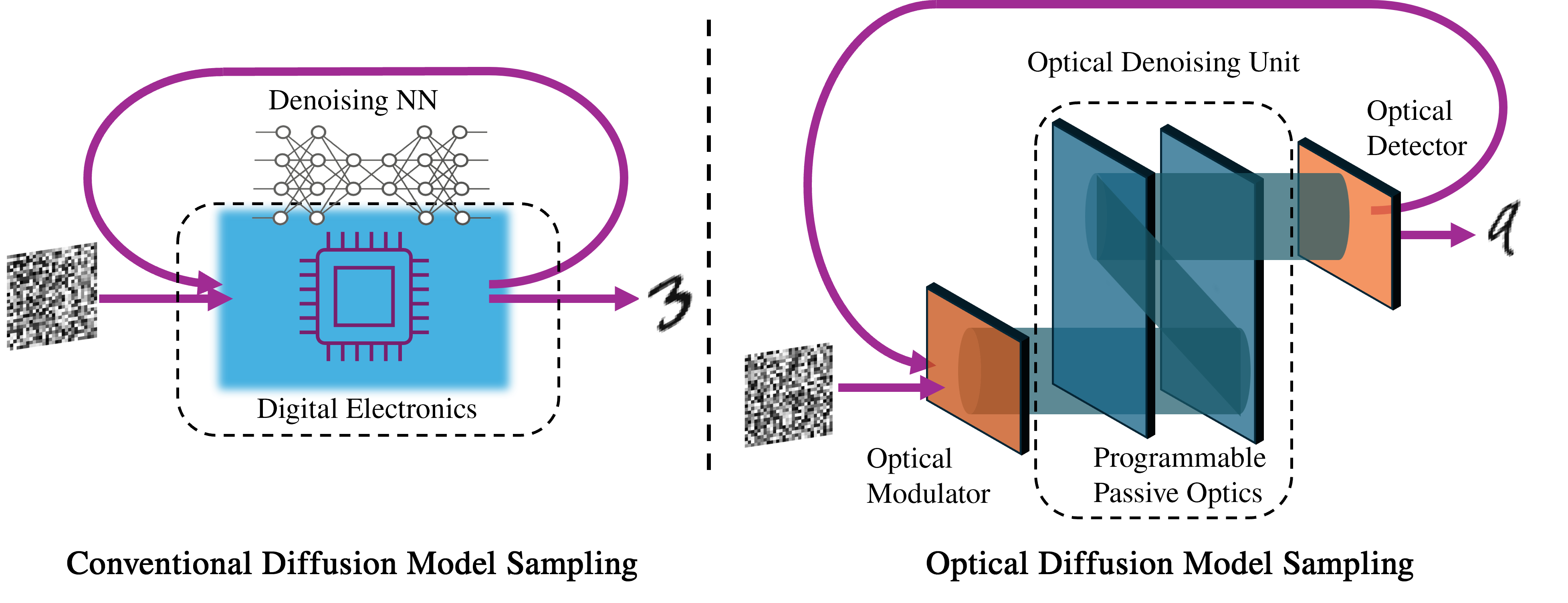}
  \medskip
  \caption{Comparison between conventional and proposed methods of image generation based on diffusion models. The conventional method runs on digital electronics based computing units such as GPUs or TPUs. The proposed method utilizes an optical denoising unit that is formed by passive optical layers. The image to be denoised is sent to the system with a modulator and the output is read out with a detector.}
  \label{fig:main}
  \medskip
\end{figure}

In this study, we demonstrate that optical wave propagation can be programmed to act as a computing engine specifically designed for implementing denoising diffusion models. As light passes through specially engineered transparent layers, features related to the original distribution are filtered out without any additional power consumption or computing latency, as depicted in Figure \ref{fig:main}. This is due to the passive nature of the transparent layers, which are minimally absorptive and do not require active components or external power to function. These layers are designed to manipulate the light solely through their physical structure, which allows for the noise prediction to exit the system efficiently. Near zero energy consumption of passive optical components reduces the overall power requirements, making the system more energy-efficient.

Through iterative noise prediction and removal, the Optical Denoising Unit (ODU) can generate new images using a minimal number of these passive optical modulation layers. Since these layers do not need power or active control, they do not introduce any latency or energy overhead. Constrained only by optoelectronic input and readout hardware, this approach has the potential to significantly reduce the computational time and energy consumption of diffusion models, specifically performing inference in more sustainable and scalable ways.

The main contributions of this study are: 
\begin{itemize}
    \item The propagation of light through multiple modulation layers is programmed to perform denoising diffusion image generation by predicting and transmitting the noise term in the input images. The system uses only a single modulation plane and multiple reflections. 
    
    \item A time-aware denoising policy is specifically designed for analog optical computing hardware. This policy facilitates the use of passive building blocks to achieve multi-step computing at low power, translating the time-embedding in digital Denoising Diffusion Probabilistic Models (DDPMs) into optical hardware.
    
    \item An online learning algorithm is introduced for training ODUs in real-life scenarios, where alignment and calibration errors exist. The algorithm tracks and alleviates experimental discrepancies with constant updates to a digital twin (DT) during training time.
    
\end{itemize}

\section{Related Work}
    Diffusion models have become popular, with their superior image generation performance compared to  Generative Adversarial Networks (GANs) \cite{dhariwal_diffusion_2021}. High-resolution, guided diffusion process is currently widely utilized for on-demand image generation \cite{saharia_photorealistic_2022,rombach_high-resolution_2022,ramesh_hierarchical_2022}. One of the main concerns with these highly capable models is the significant time taken for generating new samples, which can exceed 10 seconds for each high-resolution image \cite{rombach_high-resolution_2022}. Different methods have been proposed to alleviate this condition and improve the efficiency of diffusion models. While Latent Diffusion Models \cite{rombach_high-resolution_2022} work on a lower dimensional representation of images to decrease the computational load, Denoising Diffusion Implicit Models \cite{song_denoising_2020} introduce a deterministic and non-Markovian sampling process to reduce the number of required steps. Similarly, FastDPM \cite{kong_fast_2021}, uses domain-specific conditional information for faster sampling with diffusion models. Another approach is to distill multiple denoising steps to a single one with a teacher-student setting \cite{salimans_progressive_2022}. As these methods aim to decrease iterative denoising steps required for sampling through algorithmic innovations, the potential improvements obtained by exploiting the repetitive nature of these models on the computing hardware side remain to be seen.

Optical processors have shown substantial energy efficiency improvements, particularly with larger model sizes, potentially outperforming current digital systems \cite{anderson_optical_2023}. They can be implemented through various architectures, each leveraging different aspects of optical technology to perform computations. Free-space optical networks use spatial light modulators or fixed modulation layers for performing matrix multiplications and convolutions as light propagates \cite{zhou_large-scale_2021, lin_all-optical_2018,hu_diffractive_2024}, making them highly efficient for image processing tasks. These applications include super-resolution \cite{isil_super-resolution_2022}, noise removal \cite{isil_all-optical_2024}, and implementation of convolutional neural network layers \cite{chang_hybrid_2018}, which are shown to competitively perform different downstream tasks such as segmentation \cite{wei2023spatially}. On the other hand, photonic integrated circuits utilize optical components like Mach-Zehnder interferometers, microring resonators, and waveguides on a single chip, enabling compact vector-matrix multipliers \cite{shen_deep_2017,tait_silicon_2019}. Together, these developments highlight the transformative potential of optical computing in enhancing the performance and efficiency of computationally intensive tasks.

Considering the significant computational demands of denoising tasks, there is a clear need for specialized hardware to scale these operations effectively. Despite the advancements in optics, deep learning, and optical image processing, the realization of an optical diffusion denoiser remains a gap in current research. Bridging this gap could leverage the synergy between these fields to develop highly efficient and scalable solutions for denoising diffusion models.

In addition to their wide range of advantages, optical computing systems also have disadvantages related to the energy cost of modulation and detection of light, its limited programmability, and experimental precision. Calculating the gradients of the experimental loss through the optical wave propagation model allows for a close match between optical experiments and computational models \cite{zhou_large-scale_2021}. Similarly, a pre-trained neural network-based emulator of a physical system can also be used for the same purpose \cite{wright_deep_2022}. Moreover, it is crucial to perform as many computations as possible with the data while in the optical domain to avoid energy and time expenditure of optoelectronic devices.
 
\section{Description of the Study}

\subsection{Denoising Diffusion Models}
DDPMs progressively corrupt data with Gaussian noise in a forward process and subsequently learn to reverse this corruption through a denoising process. This way they can generate new data samples that closely resemble the training data distribution. The forward diffusion process involves the sequential corruption of a data sample $r_0 \sim q(r_0)$ through the addition of Gaussian noise over $T$ timesteps. At each timestep $t$, the data sample $r_{t-1}$ is perturbed to produce $r_t$, $r_t = \sqrt{1 - \beta_t} r_{t-1} + \sqrt{\beta_t} \epsilon_t$, where $\beta_t \in (0, 1)$ is a variance schedule that determines the amount of noise added and $\epsilon_t \sim \mathcal{N}(0, \mathbf{I})$ is standard Gaussian noise. This process transforms the original data into nearly pure noise by timestep $T$.

The reverse denoising process in DDPMs aims to reconstruct the original data from a highly noisy
sample. Starting from completely Gaussian noise $r_T \sim \mathcal{N}(0, \mathbf{I})$, the sample is iteratively denoised by removing the prediction of $\epsilon_t$ in the image,  $\epsilon_\theta(r_t, t)$, which is provided by a trained neural network: 
\begin{equation}
r_{t-1} = \frac{1}{\sqrt{1 - \beta_t}} (r_t - \beta_t \epsilon_\theta(r_t, t)),
\end{equation}

The training objective of the neural network can be simplified to minimize the mean squared error (MSE) between the true noise $\epsilon_t$ and the predicted noise $\epsilon_\theta(r_t, t), 
\mathcal{L} = \mathbb{E}_{t, r_0, \epsilon_t} \left[ \|\epsilon_t - \epsilon_\theta(r_t, t)\|^2 \right]$
where $t$ is uniformly sampled from $\{1, \ldots, T\}$. Finally, to generate new data samples, the model starts with a sample $r_T \sim \mathcal{N}(0, \mathbf{I})$ and applies the learned reverse transitions iteratively.

\subsection{Propagation of Modulated Light Beams}\label{propagation}

\begin{figure}[!htbp]
  \centering
  \includegraphics[width = 1.0\textwidth]{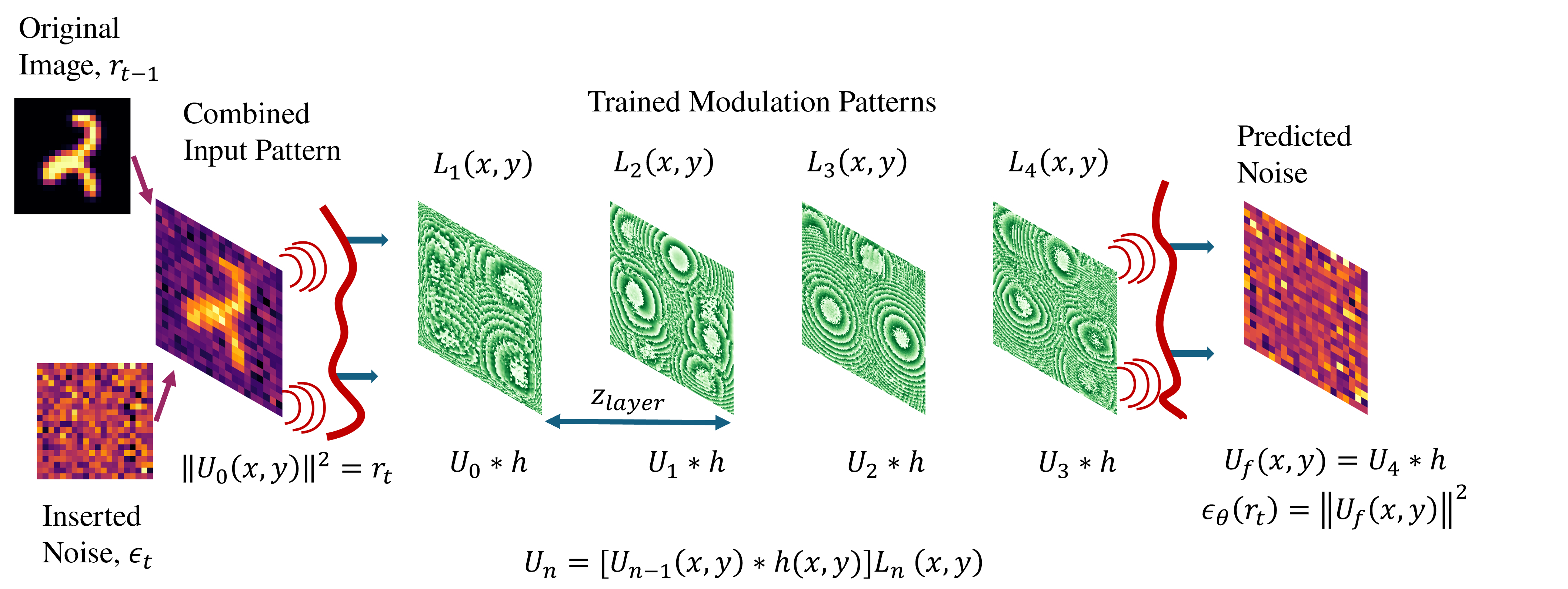}
  \medskip
  \caption{The main operation principle of ODU. Consequent modulation and free space propagation events can be represented with multiplication and convolution operations. When the input beam $U_0(x,y)$, which is patterned with noisy input images, $r_t$, is introduced to the ODU, the output intensity pattern $\|U_f(x,y)\|^2$ corresponds to the trained optical system's prediction of the noise component in the input pattern, $\epsilon_\theta(r_t)$. }
  \label{explanation_figure}
  \medskip

\end{figure}

In this study, a denoising framework is presented by combining the modulation of a light beam with consequent transparent or reflective patterns and its propagation in free space (environments such as vacuum or air, where the refractive index of light is approximately 1), as shown in Figure \ref{explanation_figure}. This process can be explained by the Fresnel diffraction theory since the features on the layers are not only larger than the optical wavelengths but only sufficiently smaller than the distance between different modulation layers \cite{goodman_introduction_2017}. According to this formalism, the electromagnetic field after propagating a distance $z$ in free space, $U(x, y, z)$, can be calculated from its distribution at $z=0$ by convolution with "the impulse response of free space", $h(x, y)$:
\begin{equation}\label{fs-impulse-func}
U(x, y, z)=U(x, y, 0) * h(x, y), \text{ where } h(x, y)=\frac{e^{j k z}}{j \lambda z} \exp \left[\frac{j k}{2 z}\left(x^2+y^2\right)\right]
\end{equation}
Here, $k$ denotes the wavenumber of the field, and $\lambda$ is the wavelength. In other words, the field's value at the plane of $z=z_0$, at a given location $(x,y)$, is the weighted sum of the values at $z=0$, where the weight of each location is determined by the response function. Being complex numbers, all of these weights have the same magnitude but their phase depends on the location. 
In the frequency domain, the transfer function of free space becomes
\begin{equation}
H\left(f_X, f_Y\right)=\exp \left[j 2 \pi \frac{z}{\lambda} \sqrt{1-\left(\lambda f_X\right)^2-\left(\lambda f_Y\right)^2}\right]. \label{transfer-func}
\end{equation}
This indicates that for spatial frequencies larger than $1/\lambda$, the magnitude of the transfer function decays to zero exponentially. Hence, only features that are larger than the wavelength of the light can propagate to the far field. Moreover, frequency domain expression of diffraction, Eqn. \ref{transfer-func}, allows for also the efficient digital simulation of the propagation of light in free space with the utilization of Fast Fourier Transforms (FFTs) in a parallelized manner. Later on, we will benefit from this fact for GPU accelerated training of the diffractive modulation layers.

The proposed method applies trainable weights to the light beam at consequent planes with thin modulation layers. The interaction between layers and light can be represented as a point-wise multiplication between the incident field and the layer, which is followed by the propagation of the field in free space until the next layer,

\begin{equation}
U_n(x,y) = [U_{n-1}(x, y) * h(x, y)]L_n(x, y),
\end{equation}

where $U_n(x,y)$ is the field distribution right before reaching the modulation layer $n$, and $L_n(x, y)$ is the complex modulation coefficient distribution of the trained modulation layers or the weights of the optical diffusion model. $L(x, y) = |L(x, y)| e^{i \phi(x, y)}$ can be only a real number($\phi(x, y) = 0 $), just phase modulation ($|L(x, y)|=1$) or an arbitrary complex number depending on the implemented modulation principle. In this paper, we demonstrate our approach with a phase-only liquid crystal spatial light modulator (SLM), which can set $\phi(x, y)$ to any value in the range of $[0,2\pi]$ electronically, and $|L(x, y)| \approx 1$ everywhere.

\subsection{Training of Optical Modulation Layers}

As described in Section \ref{propagation}, propagation of light can be analytically explained in a succinct manner for the scale considered in this study ($z_{layer}>>d_{pixel}>\lambda$). This allows for defining some free variables in this representation, such as refractive index distribution $L(x,y)$ or input wavefront distribution $U_0(x,y)$, and optimizing these variables for minimizing a cost function. The gradients of these variables can be found with either manual calculation \cite{kamilov_learning_2015} or automatic differentiation packages \cite{lin_all-optical_2018}. In this study, our first goal is to find optimal modulation layers, or refractive index distributions, such that after the light beam encoded with the noisy images propagates through them only the predicted noise term reaches the detector, as shown in Figure \ref{explanation_figure}. Moreover, the denoising network should be aware of the given timestep in the diffusion process while predicting the noise, $\epsilon_\theta(x_t, t)$, so that it would have \textit{a priori} information about the variance of the noise term. Since noise level awareness is a crucial aspect of successful sample generation, most of the current implementations of diffusion models utilize time-embedding layers to modify activations of the neural network across different layers depending on the diffusion time step. Instead, the proposed method divides the diffusion timeline consisting of $T$ timesteps into $M$ subsets, and for each subset of time frames \( \{S_m\}_{m=1}^M \), trains a separate set of modulation layers $\{L_n^m\}_{n=1}^N$ each containing $N$ layers. Then, for \( t \in S_m \), the noise prediction,$\epsilon_{\theta_m}(x)$ becomes only a function of $x$. In this scheme, the training objective for each time step is
\begin{equation}
\mathcal{L}_t = \mathbb{E}_{x_0, \epsilon} \left[ \left\| \epsilon - \epsilon_{\theta_m}(x_t) \right\|^2 \right]
\end{equation}
where  total loss is the sum over all ranges:
\begin{equation}\label{loss-func}
\mathcal{L} = \sum_{i=m}^M \sum_{t \in S_m} \mathcal{L}_t.
\end{equation}

This decoupling of denoising at different timesteps by removing time-embedding layers also eliminates the necessity for digital computations to modifications at different layers. By circumventing this problem we perform denoising all-optically. Moreover, a fixed optical modulation pattern performs denoising at multiple consequent timesteps. For instance, we later demonstrate that for $T=1000, M =10$ creates optimal results. So, a single layer set can process 100 timesteps and the entire sampling workflow can be operated with only 10 fixed parallel devices, or with only 10 updates to the SLM.

After defining the forward calculation of the system with the analytical explanation of light propagation and the loss function as the mean square error of noise prediction (Eqn. \ref{loss-func}), the trainable parameters of the system $\{L_n^m\}_{n=1}^N (x,y)$ are optimized by automatic differentiation \cite{paszke_automatic_2017}.

\section{Results}

\begin{figure}[!h]
  \centering
  \includegraphics[width = 0.8\textwidth]{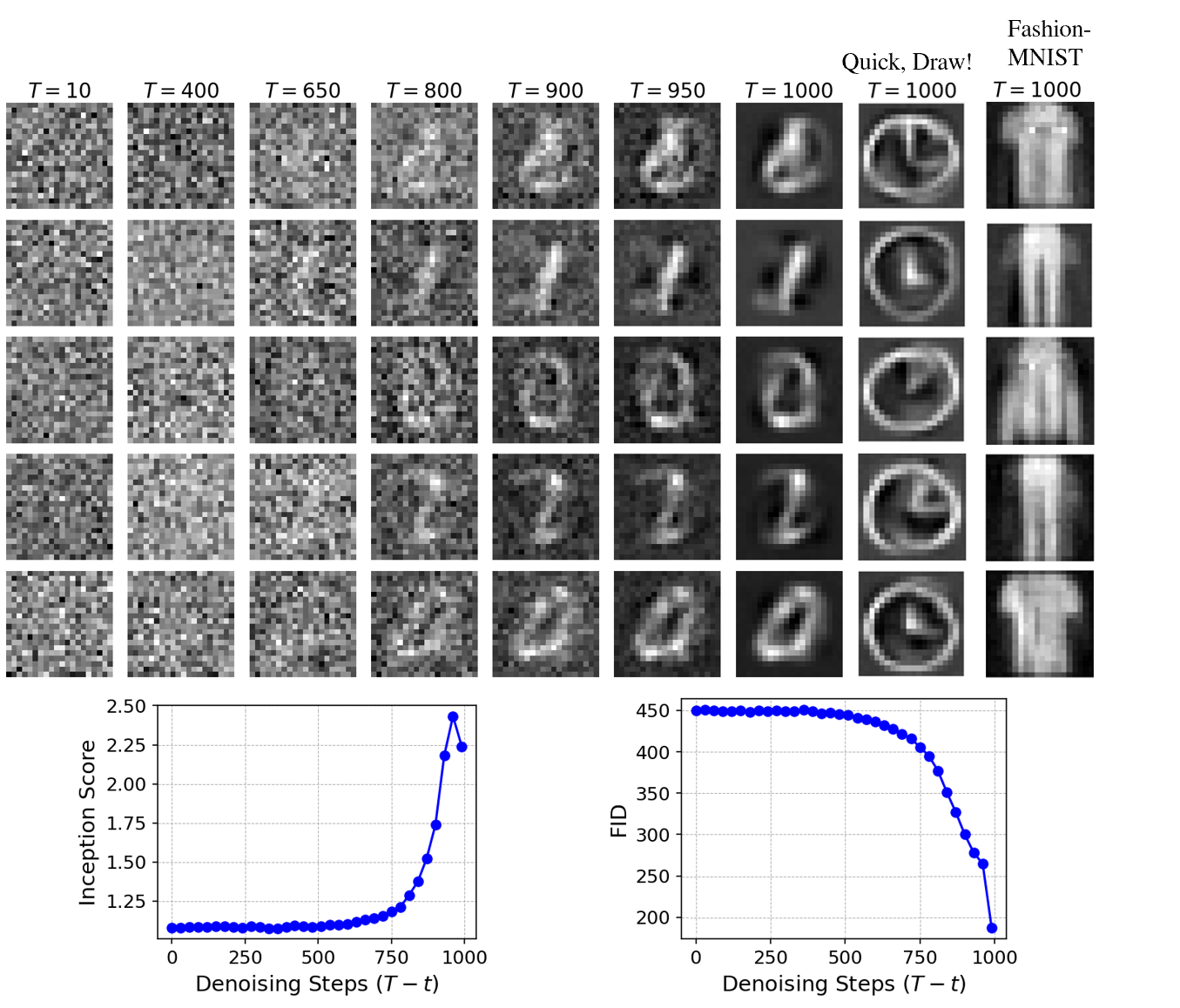}
  \caption{Images generated by the Optical Diffusion Model at different timesteps and when trained with various datasets. The generated images and their corresponding Inception and FID scores are calculated between timesteps $T = 10$ to $T=950$ are acquired after training with the MNIST digits dataset. Final outputs at time $T=1000$, acquired from ODUs trained for the MNIST digits samples have FID = $206.6$, for Fashion MNIST,  FID = $227.7$ and for Quick, Draw!, FID = $131.4$}
  \label{denoise-results}
  \medskip

\end{figure}

\subsection{All-Optical Denoising based Image Generation}\label{sec:image-generation}

Following the same experimental settings with the initial DDPM study \cite{ho_denoising_2020}, we set $T = 1000$ and $\beta$ values to be in the linear  range between $\beta_1 = 10^{-4}$ to $\beta_T = 0.02$. The results in Figure \ref{denoise-results} are reported with the beam propagation model (Eqn. \ref{transfer-func}) of the optical system designed to have $300\times300$ pixels per layer and four modulation layers. The number of layer sets ($M$) is 10. Several intermediate results alongside final outputs at $T = 1000$ are reported in Figure \ref{denoise-results} for 3 classes of the MNIST digits \cite{lecun1998gradient}, Fashion-MNIST \cite{xiao2017/online} and the clock category of the Quick, Draw! datasets \cite{noauthor_googlecreativelabquickdraw-dataset_2024}. Furthermore, the evaluation of image generation quality metrics, Inception Score (IS) and Fréchet Inception Distance (FID), which are detailed in Appendix \ref{appendix:metric-definition}, across different generation timesteps captures the improved realism of images with the optical diffusion procedure.

\subsection{Effects of Optical Model's Dimensionality on the Image Denoising and Generation Performance} \label{sec4-2}

\begin{figure}[h]
  \centering
  \includegraphics[width = 0.8\textwidth]{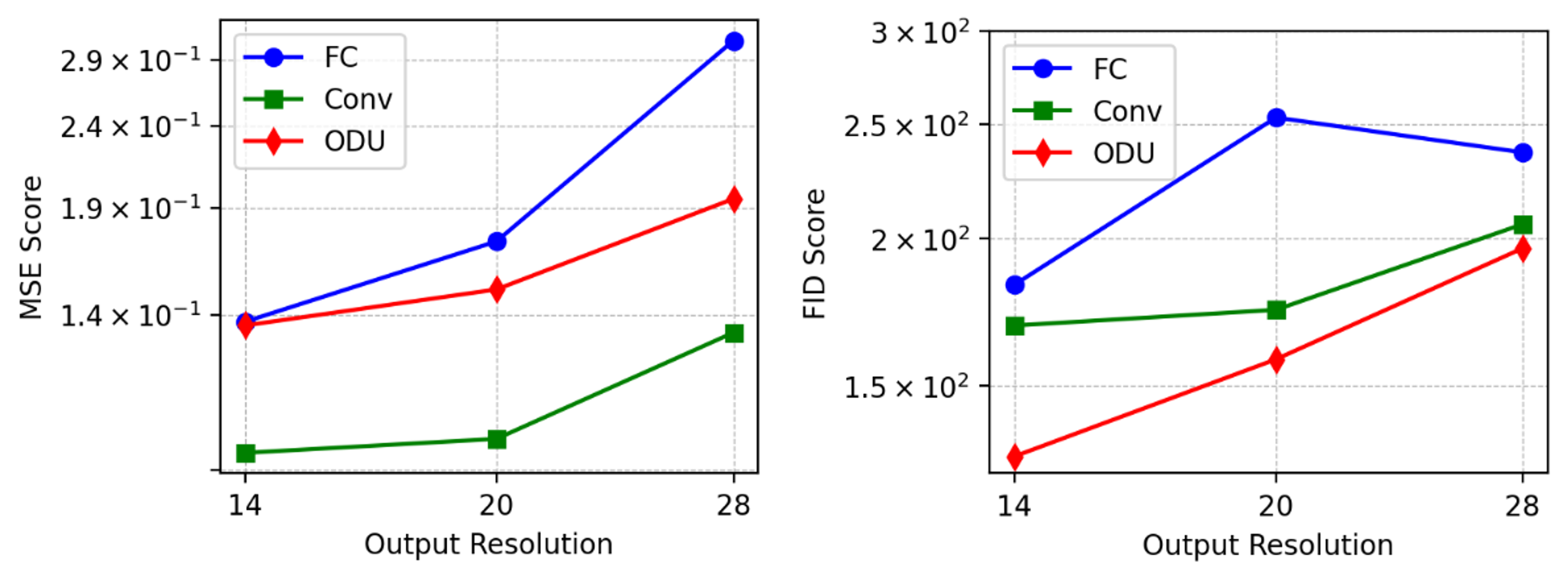}
  \medskip
  \caption{Scaling of the denoising capabilities (left) and generation performance (right) of Optical Diffusion, and pure digital convolutional U-Net and fully connected networks with the output image resolution.}
  \label{fig:output-res-scaling}

\end{figure}

This section provides further analysis with different output dimensions and parameter counts along with the comparisons with purely digital implementations to quantify Optical Diffusion's scalability to large-scale diffusion problems. The first investigation is into the performance with higher resolution datasets; two digital architectures of similar performance with the ODU, one being fully connected and the other convolutional U-Net \cite{ronneberger2015u}, are trained for the same tasks with the ODU, generating images with the MNIST digits dataset at different resolutions. Their architecture is detailed in Appendix Table \ref{tab:architecture_performance}. The results shown in Figure \ref{fig:output-res-scaling} indicate that ODU consistently outperforms the two digital neural networks, and all three scale in a similar manner both in terms of denoising and generation performances when the generated image dimension is increased while the model sizes are kept constant. 

\begin{figure}[h]
  \centering
  \includegraphics[width = 0.95\textwidth]{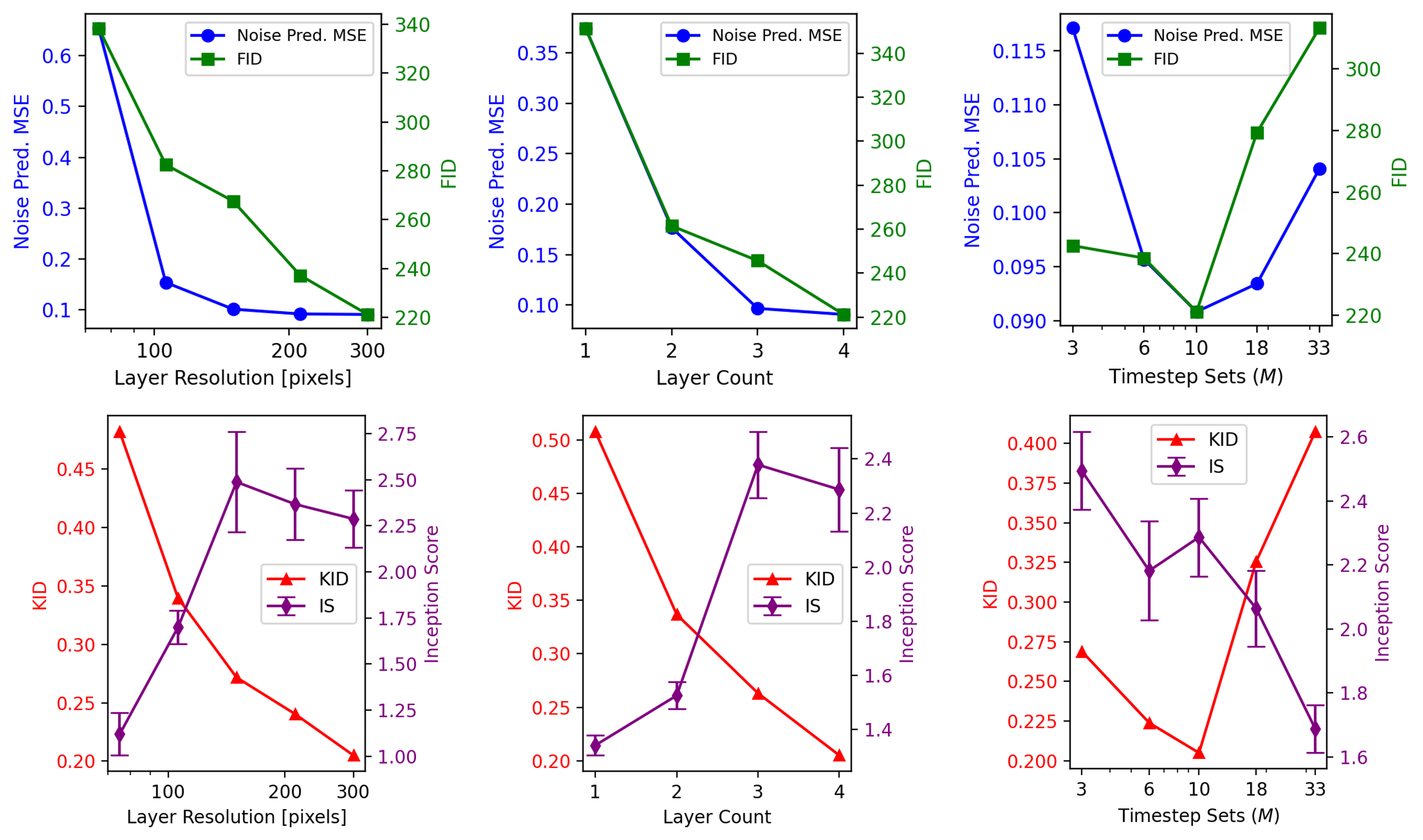}
  \medskip
  \caption{The dependency of denoising performance (MSE) and generation quality scores(FID, KID and Inception score), on the hyperparameters of the ODUs (number of pixels of optical modulation layers, number of modulation layers and number of denoising layer sets ($M$)). }
  \label{timestep}
  \medskip

\end{figure}

Secondly, the scaling of Optical Diffusion's performance with respect to the number of total parameters is probed through three hyperparameters: layer resolution, layer count, and timestep sets. The effects are tracked with different metrics, MSE for denoising, FID, IS and Kernel Inception Distance (KID) for the generation quality. In Figure \ref{timestep}, we observe that, as in digital neural networks, there is a clear tendency to perform better with a larger number of trainable parameters when layer resolution and layer count are increased.

On the other hand, having a larger number of denoising layer/timestep sets improves the results only until they reach a certain level. Afterward, increasing the number of sets is detrimental as shown in Figure \ref{timestep}. As the total number of training steps is fixed in this experiment, increasing the number of timestep sets decreases the training sample count per layer set, hence potentially deteriorating the performance after a particular threshold, which is found to be $M=10$.


\begin{figure}[!h]
  \centering
  \includegraphics[width=0.50\textwidth]{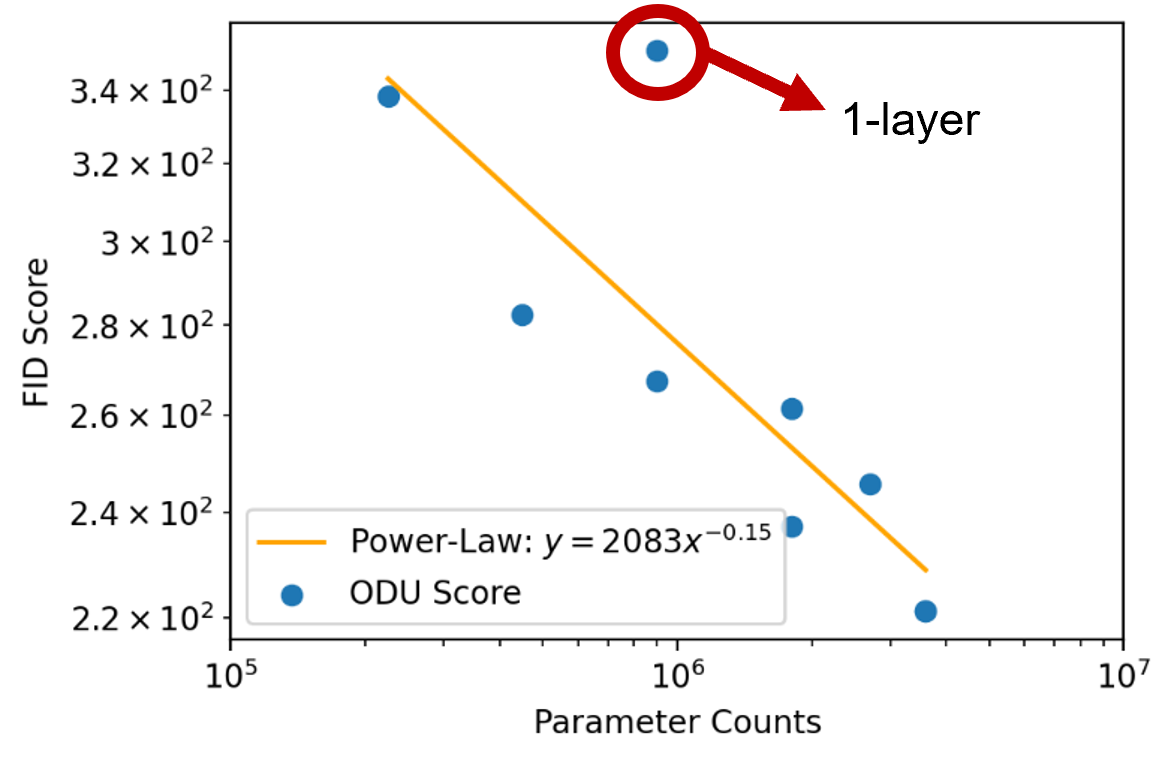} 
  \medskip
  \caption{The relationship between the total number of parameters in an ODU and its generation performance in terms of FID scores.}
  \label{fig:total-params}
  \medskip

\end{figure}

Through the aggregation of data points acquired with different layer resolutions and counts in Figure \ref{timestep}, the relationship between the total trainable parameter count of ODUs and their image generation performance is depicted in Figure \ref{fig:total-params}. This relationship remarkably follows the same widely accepted, power-law trend of digital generative models \cite{henighan2020scaling}. Most significantly, when the optical implementation is fitted to a power-law equation, the exponential of the power law ($-0.15$) is approximately the same as the reported value ($-0.16$) for large-scale image generation networks in \cite{henighan2020scaling}. This fit parameter gives the slope of the line in the logarithmic plot, indicating how fast the generation performance scales with the number of parameters, in this case showing that ODU improves its performance at a similar speed with large-scale digital image generation networks while its parameter count is increased. The single outlier in this trend is the case where there is only a single modulation layer, which does not benefit from the multiple optical modulations aspect of the proposed architecture.

\subsection{Higher Experimental Fidelity with the Online Learning Algorithm} \label{sec:online-learning}

 To address the challenge of training an optical system with imperfect calibration, as faced in many other analog computing paradigms, we propose an online learning algorithm that updates and leverages a DT during training. During inference, the DT does not incur any additional overhead. The DT ($\tilde{f}_{\theta_{\text{layers}}, \theta_{\text{alignment}}}$) again utilizes Fresnel diffraction based model of light propagation, as a surrogate to compute gradients and guide the optimization of the system's trainable parameters. However, matching the DT's parameters ($\theta_{\text{alignment}}$), for instance, input beam angle, precise locations of the layers, and their angles, perfectly to the experimental conditions of the physical system is a challenging task. Therefore, during each iteration of training, the output of the experiment ($f_{\theta_{\text{layers}}}$) and DT ($\tilde{f}_{\theta_{\text{layers}}, \theta_{\text{alignment}}}$) is compared and the DT's parameters are updated accordingly, as shown in Algorithm \ref{pseudo-algo}.

\begin{algorithm}
\caption{Online Learning Algorithm}
\label{pseudo-algo}
\begin{algorithmic}[!htbp]
\State Initialize physical system \(f_{\theta_{\text{layers}}}\) with parameters \(\theta_{\text{layers}}\)
\State Initialize DT \(\tilde{f}_{\theta_{\text{layers}}, \theta_{\text{alignment}}}\) with parameters \(\theta_{\text{layers}}, \theta_{\text{alignment}}\)
\While{not converged}
    \State \textbf{Forward Pass:}
    \State Input data \(\mathbf{x}\) into the physical system \(f_{\theta_{\text{layers}}}\)
    \State Obtain physical system output \(\mathbf{y}_{f}\) = $f_{\theta_{\text{layers}}}(\mathbf{x})$ 
    \State Compute error \(E = \text{loss}(\mathbf{y}_{f}, \mathbf{y}_{\text{target}})\)
    
    \State \textbf{Backward Pass:}

    \State Compute Jacobian of DT at  \(\mathbf{x},\mathbf{J} = \frac{\partial \mathbf{y}_{\tilde{f}}}{\partial \theta_{\text{layers}}}\)
    \State Compute gradients \(\nabla_{\theta_{\text{layers}}} = \mathbf{J}^T \cdot \frac{\partial E}{\partial \mathbf{y}_{f}}\)
    \State Update physical system parameters \(\theta_{\text{layers}} \leftarrow \theta_{\text{layers}} - \eta \nabla_{\theta_{\text{layers}}}\)

    \State \textbf{DT Refinement:}
          \State Obtain DT output \(\mathbf{y}_{\tilde{f}}\)

    \State Compute MSE between DT and physical system outputs \(L = \text{MSE}(\mathbf{y}_{\tilde{f}}, \mathbf{y}_{f})\)
    \State Compute gradients \(\nabla_{\theta_{\text{alignment}}} = \frac{\partial L}{\partial \theta_{\text{alignment}}}\)
    \State Update DT alignment parameters \(\theta_{\text{alignment}} \leftarrow \theta_{\text{alignment}} - \alpha \nabla_{\theta_{\text{alignment}}}\)
\EndWhile
\end{algorithmic}
\end{algorithm}

In parallel, the DT is also employed to compute the gradients of the trainable parameters of the experiment, with respect to the output($\frac{\partial \mathbf{y}_{\tilde{f}}}{\partial \theta_{\text{layers}}}$). These gradients are then utilized to update the physical system's parameters through backpropagation, informed by the error obtained from the physical system. Concurrently, the DT is refined using the latest inputs and outputs from the physical system to better approximate its behavior, despite the initial parameter mismatches. This iterative process of forward and backward passes, coupled with the continuous refinement of the DT, enables the physical system to progressively improve its performance and align more closely with the desired outcomes.

\begin{figure}[!htp]
  \centering
  \includegraphics[width = 0.8\textwidth]{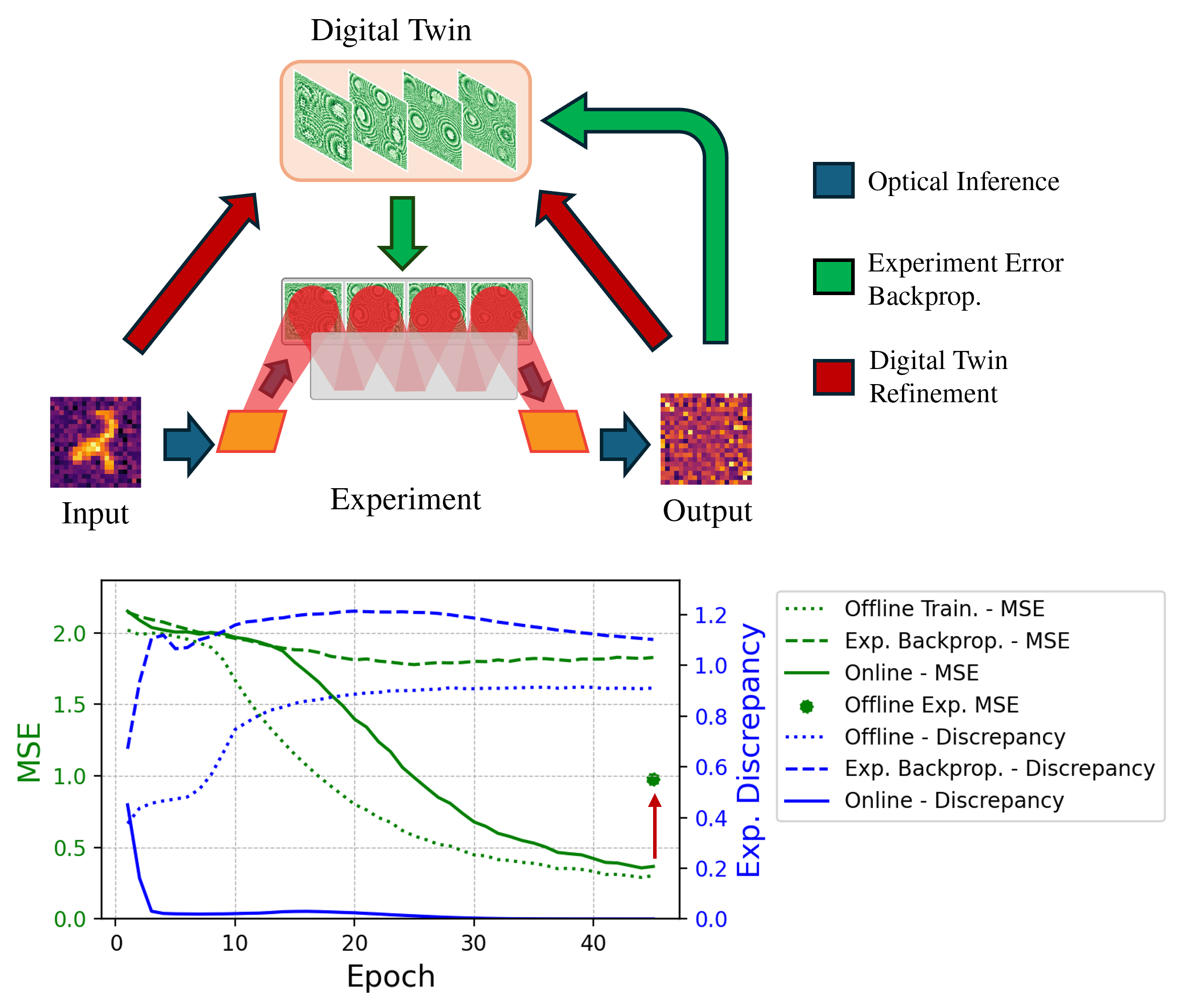}
  \medskip
  \caption{The upper block illustrates the online training scheme. The forward pass is calculated with the experiment (blue), while the gradients of the prediction error are backpropagated using a DT of the experiment (green) and updating the physical trainable parameters. The difference between the outputs of the experimental setup and the DT continuously refines the DT (red). The lower graph block compares offline and online training methods. Offline training relies on a pre-trained DT for both the forward and backward passes, with the experimental performance of this method indicated by the star. Experimental backpropagation executes the physical forward pass but does not incorporate DT refinement.}\label{online-learning}
  \label{exp-online}
  \medskip

\end{figure}

In Figure \ref{online-learning}, we explore the efficiency of the proposed algorithm by modeling a possible experiment scenario.  In this scenario, we define two different optical models, while actually both of them are simulations of the optical wave propagation for an exact insight into the algorithm, we designate the first one as the “optical experiment” by configuring it with the calibration angles obtained from the physical experiment.  These four angles account for the slight misalignment of the experiment and define the input angle of the beam to the cavity and the angle between the mirror and the SLM, in x and y axes, all being in the range of a few milliradians and their measurement details being provided in Appendix \ref{differentiable-modelling}. The second model, considered as the DT, is initialized with their calibration angles $20\%$ higher.
We used 3 different algorithms, offline, experimental error backpropagation \cite{wright_deep_2022}, and the proposed online training schemes. During training, MSE (training loss), and the discrepancy between the DT and the experiment's outputs are tracked. The discrepancy, $D$, is inversely related to cross-correlation, $C$, of the experimental and the DT's normalized outputs, $O_{exp}$ and $O_{dt}$ respectively, $D = 1-C$, where $C = \sum_{x} \sum_{y} O_{\text{exp}}(x, y) O_{\text{dt}}(x, y)$.

Offline training improves the loss function when evaluated with the DT, but when evaluated in the experimental setting, it has a higher MSE, as shown with a star in Figure \ref{online-learning}. When the online training method is used, the DT is aligned with the actual experiment swiftly and the experimental loss approximates the MSE of perfect calibration case, the results of this approach with the optical experiment are also provided in Appendix Figure \ref{fig:experimental_system_results}. Backpropagating the experimental loss, MSE does not decrease significantly. However, for smaller misalignment, this method was also demonstrated to converge. This experiment implements multiple modulation layers on a single device, with a single phase-only SLM and a mirror in parallel to resource-efficiently prototype the proposed computing method, as shown in Figure \ref{exp-online} and detailed in Appendix \ref{details-experimental}.

\section{Conclusion}

In this study, we introduced an optical diffusion denoising framework for image generation, utilizing a time-aware denoising strategy that enables optical low-power realization. By exploiting light propagation through transparent media, ODU effectively reduces noise in images, with a much smaller energy budget compared to electronics since optical wave propagation has a very small intrinsic loss while acquiring comparable, or better quality. This is especially interesting because diffusion models are currently one of the most costly generative artificial intelligence models due to their repetitive denoising process, with a correspondingly large environmental impact \cite{10.1145/3630106.3658542}. 

The integration of a time-aware policy enables Optical Diffusion to adjust light modulation dynamically according to different stages of the denoising process, thus improving image quality with a minimal number of changes to the modulation layers or parallel optical processing units. Looking ahead, the incorporation of larger modulation layers with more parameters and the exploration of nonlinear optical effects could enhance the functionality of the system. Scaling analyses also show evidence for the ODU to improve its performance at the same rate as digital models while increasing its size. These potential improvements suggest promising directions for future research in expanding the range of applications for this technique.

The proposed method can utilize off-the-shelf consumer electronics such as digital micromirror devices that can be found in portable projectors for input modulation and CMOS cameras for recording output prediction. The online learning algorithm accounts for variations between these devices and closes the gap between the analytical and experimental realization of the ODU. On the other hand, as analyzed in detail in Appendix \ref{speed-energy}, these devices have the potential of implementing denoising steps on the order of microsecond latencies while consuming a few Watts only.
With the utilization of high-speed light modulation technologies, million-frames-per-second-level denoising can be achieved again with Watt level energy consumption, while still utilizing the proposed approach for predicting noise in provided images \cite{panuski_full_2022}.

\begin{ack}
This work was supported by Google Research under grant number 901381. We would like to thank Research Computing Platform at EPFL and Google Cloud Services for providing access to compute units.
\end{ack}

\medskip
\clearpage
\appendix

\section{Appendix}
\subsection{Differentiable Modelling of Light Propagation in ODUs}\label{differentiable-modelling}
To benefit from parallelized and optimized FFT algorithm and automatic differentiation, the wave propagation in the proposed system is modeled in PyTorch environment with a Split-Step Fourier formalism which is derived from Eqn. \ref{fs-impulse-func}. The diffraction step of the propagation is calculated with a nonparaxial diffraction kernel \cite{feit1988beam} in Fourier domain and effects such as reflection or modulation of the light beam with layer parameters are applied in the spatial domain. Such that the electric field after propagating a distance $\Delta z$ becomes:
\begin{equation}
    E(x, y, z+\Delta z) =  \mathcal{F}^{-1} \{ \mathcal{F}\{ E(x,y,z) R(x,y)  \}e^{-\frac{j \Delta z \left(k_x^2 + k_y^2\right)}{k +\sqrt{k^2 - k_x^2 - k_y^2}}}\}
\end{equation}

In addition to the parameters of layers $L_n^m (x, y)$ (or simply $\theta_{layers}$), the spatial term $R(x,y)$ can include the angle changes of the beam. For instance, if the beam is not perpendicular to the SLM or the mirror, the reflection creates a change in the angle of the beam, $\Delta \alpha = (\alpha_x, \alpha_y)$. Then, on the SLM plane $R_m(x, y)=L_m(x, y) e^{-j k\left(x \sin \alpha_x+y \sin \alpha_y\right)}$, where $R_m(x,y)$ is the compound spatial term, $L_m(x,y)$ is the modulation parameters at layer $m$, and $e^{-jk(x \sin \alpha_x + y \sin \alpha_y)}$ is the operator that changes the direction of the wave propagation vector. Similarly to the SLM, also the angle of the mirror determines the propagation direction of the beam, which can be included in the model as $R_{\text {mirror }}(x, y)=e^{-j k\left(x \sin \alpha_x+y \sin \alpha_y\right)}$.
To calibrate the model of the experiment with the actual experiment, we define three trainable alignment parameters, $z_{gap}$ (distance between the mirror and the SLM), $\Delta \alpha_{mirror}$ (twice the angle between the mirror and the SLM), and $\Delta \alpha_{beam}$  (twice the angle between the input beam and the SLM). This group of trainable model parameters is called $\theta_{alignment}$  and as its constituents appear in the forward model within differentiable functions, the auto-differentiation algorithm \cite{paszke_automatic_2017} can calculate their derivatives with respect to the error between the predicted camera images and the acquired ones. $\theta_{alignment}$  is initially pre-trained with experiments placing square shaped $\pi$ phase differences randomly on the SLM. During the online training procedure, it is further trained with the data from denoising experiments.

\subsection{Details of Scaling Studies}\label{appendix:scaling}
\subsubsection{Scaling with Output Image Resolution}\label{appendix:scaling-output}

\begin{table}[!hbp]
    \centering
    \caption{Properties of Optical, Convolutional U-Net and Fully Connected Denoising Networks, the same training settings with the main text are used in all experiments with the MNIST digits dataset.}
    \begin{tabular}{@{}lcccccc@{}}
        \toprule
        \textbf{Architecture} & \textbf{Parameters} & \textbf{FLOPS/Step} & \textbf{Energy/Image [J]} & \textbf{Images/s} \\ 
        \midrule
        Fully Connected       & 19.6 M              & 39.0 M               & 1.74                      & 41.3              \\ 
        Convolutional U-Net   & 220 K               & 3.11 M               & 5.37                      & 13.9              \\ 
        ODU                   & 3.6 M               & Not Applicable       & 0.23                      & 23.0              \\ 
        \bottomrule
    \end{tabular}
    \label{tab:architecture_performance}
\end{table}

We investigated the change in the performance of Optical Diffusion when the image resolution changed while the model size was kept the same. This scaling behavior is also compared with well-established digital neural network architectures under the same diffusion settings on the MNIST digits dataset. The energy consumption and speed of the ODU in Appendix Table \ref{tab:architecture_performance} are indicated for the simple laboratory implementation where the efficiency is not optimized, while the digital benchmarks are run on an Nvidia L4 GPU, one of the state-of-the-art devices available today. 

\textbf{Fully Connected Denoising Neural Network.} This fully connected architecture consists of 4 fully connected layers with 1200 neurons, SiLU nonlinearity, and one-dimensional batch normalization layers following each fully connected layer. The outputs of the first and third layers are summed with time-embedding representations. Inputs are interpolated to $76\times76$ and flattened to vectors of 5776 elements, while the last layer outputs a same-sized vector which is again reshaped to $76\times76$ and scaled to the target resolution. During inference time, this network generated a batch of 64 images in \SI{1.55}{s}, on an NVIDIA L4 GPU utilized $100\%$ at \SI{72}{W}. This amounts to 41.3 images/s at 1.74 J/image.

\textbf{Convolutional U-Net Denoising Neural Network.} This U-Net architecture \cite{ronneberger2015u} has 2 downsampling, 1 bottleneck, and 2 upsampling blocks, featuring a total of 32 convolutional layers with $3\times3$ kernels and SiLU nonlinearity. Every block also includes time embedding and batch normalization. Similarly, inputs are interpolated to $76\times76$ pixels and the same-sized outputs are scaled to the target resolution. During inference time, this network generated a batch of 64 images in \SI{4.60}{s}, on an NVIDIA L4 GPU utilized $100\%$ at \SI{72}{W}. This amounts to 13.9 images/s at 5.37 J/image.

\textbf{ODU.} The ODUs consist of 10 sets of 4 optical layers, each with $300\times300$ modulation parameters, as detailed in Section \ref{sec:image-generation}. At an image rate of 23 kfps and total energy consumption of \SI{5.3}{W} between the DMD and the camera, the generation can be operated at 23.0 images/s at 0.23 J/image. In this scenario, the SLM is assumed to be a passive device due to the very small number of updates.

In addition to the comparison in Figure \ref{fig:output-res-scaling} using the MNIST digits dataset up to a resolution of $28\times28$, we studied further the generation quality by training with the AFHQ dataset’s cat class \cite{choi2020stargan} at $40\times40$ resolution. The results in Appendix Figure \ref{afhq-predictions} confirm the same successful scaling trend with the ODU overperforming the digital networks. Even though this relatively more complex problem necessitates larger and more capable denoising networks, similarly with the smaller scale experiments, Optical Diffusion obtained the best FID.

\begin{figure}[!htbp]
  \centering
  \includegraphics[width = 0.8\textwidth]{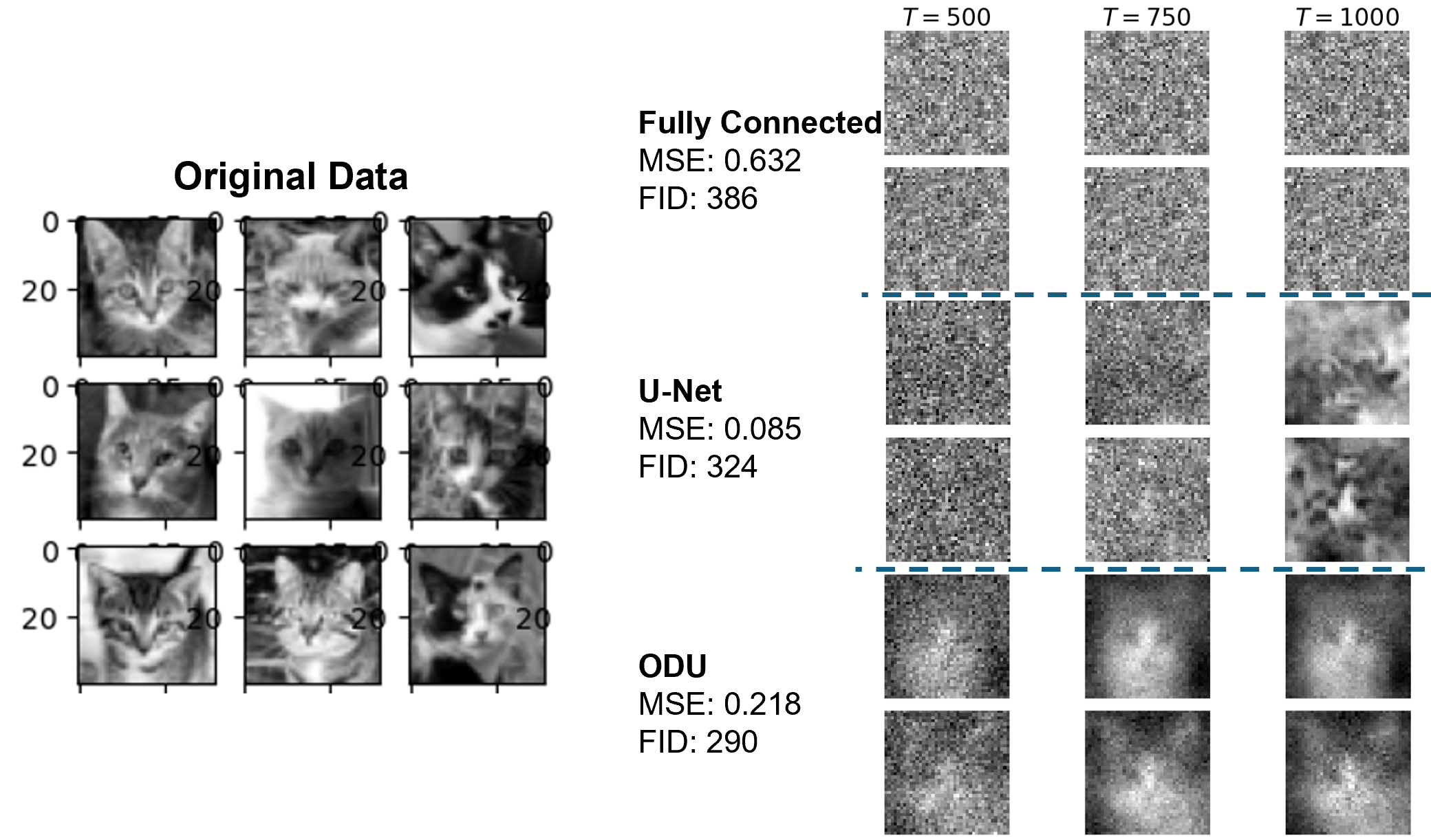}
  \medskip
  \caption{Comparison of image generation performances on the AFHQ dataset’s cat class \cite{choi2020stargan} at 40-by-40 resolution.}
  \label{afhq-predictions}
  \medskip
\end{figure}

\subsection{Details of the Experimental System and Online Learning}\label{details-experimental}

\begin{figure}[!htbp]
  \centering
  \includegraphics[width = 0.6\textwidth]{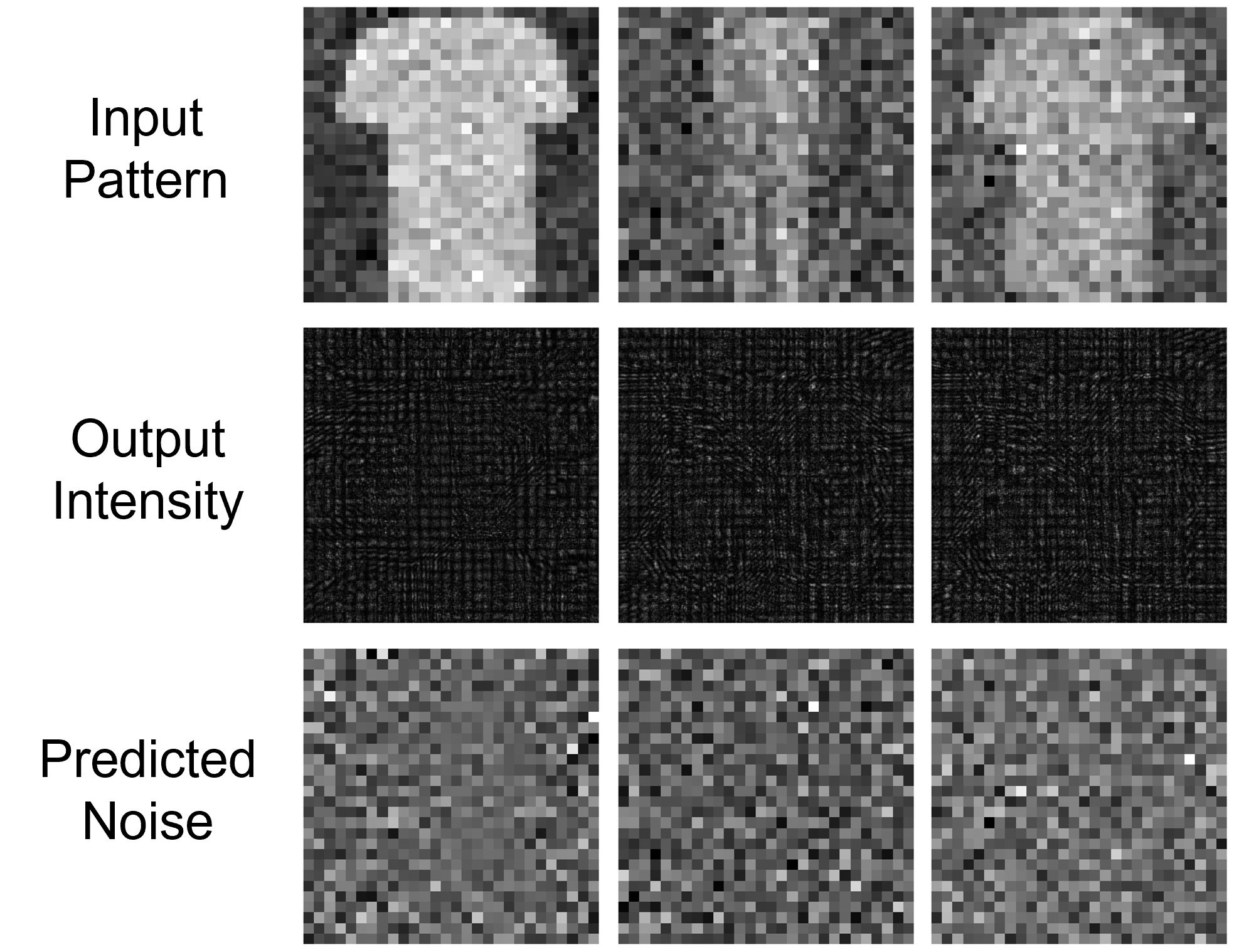}
  \medskip
  \caption{Some input patterns, output intensities at the camera plane, and the corresponding noise prediction for Fashion MNIST. The intensities on the camera plane are converted to noise predictions by downsampling and normalization.}
  \label{output-images}
  \medskip
\end{figure}

In the ODU, the trainable parameters are the pixel values on the modulation layers and are digitally adjusted using a computer model, as outlined in Appendix \ref{differentiable-modelling}. After optimizing these parameters on the computer, they are implemented across various layers in the experimental setup, utilizing the SLM. The beam generated by a continuous-wave dye laser (M-Squared Solstis 2000) at $\lambda = \SI{850}{nm}$ reaches the phase-only SLM (Meadowlark HSP 1920-500-1200) after reflecting from a digital micromirror device (DMD), which can be used for spatial amplitude modulation. After 4 reflections on the modulating area SLM, the beam is imaged onto a CMOS camera (FLIR BFS-U3-04S2M-CS). An $\SI{11.6}{mm}$  mirror, placed $\SI{17.1}{mm}$ from the SLM display, captures four reflections. To direct the input beam toward the SLM, 4F imaging was employed, transferring the beam from the DMD, which serves as a programmable aperture ensuring the beam's alignment with the first modulation layer of the SLM. We assigned $260\times260$ pixel patches for the modulation layers on the SLM, which has a pixel pitch of $\SI{9.2}{\mu m} $. After the fourth reflection, the beam is imaged with another 4F imaging system onto a CMOS camera that records the output intensity as a $130\times130$ pixel image, where the camera's pixel pitch is $\SI{3.45}{\mu m}$. The examples of input patterns to the optical system, the corresponding output intensities at the camera plane and the resulting noise pattern predictions are provided in Appendix Figure \ref{output-images}. The noise predictions are obtained by the downsampling and normalization of the pixel values in the output intensity recordings.

\begin{figure}[!h]
  \centering
  \includegraphics[width = 1.0\textwidth]{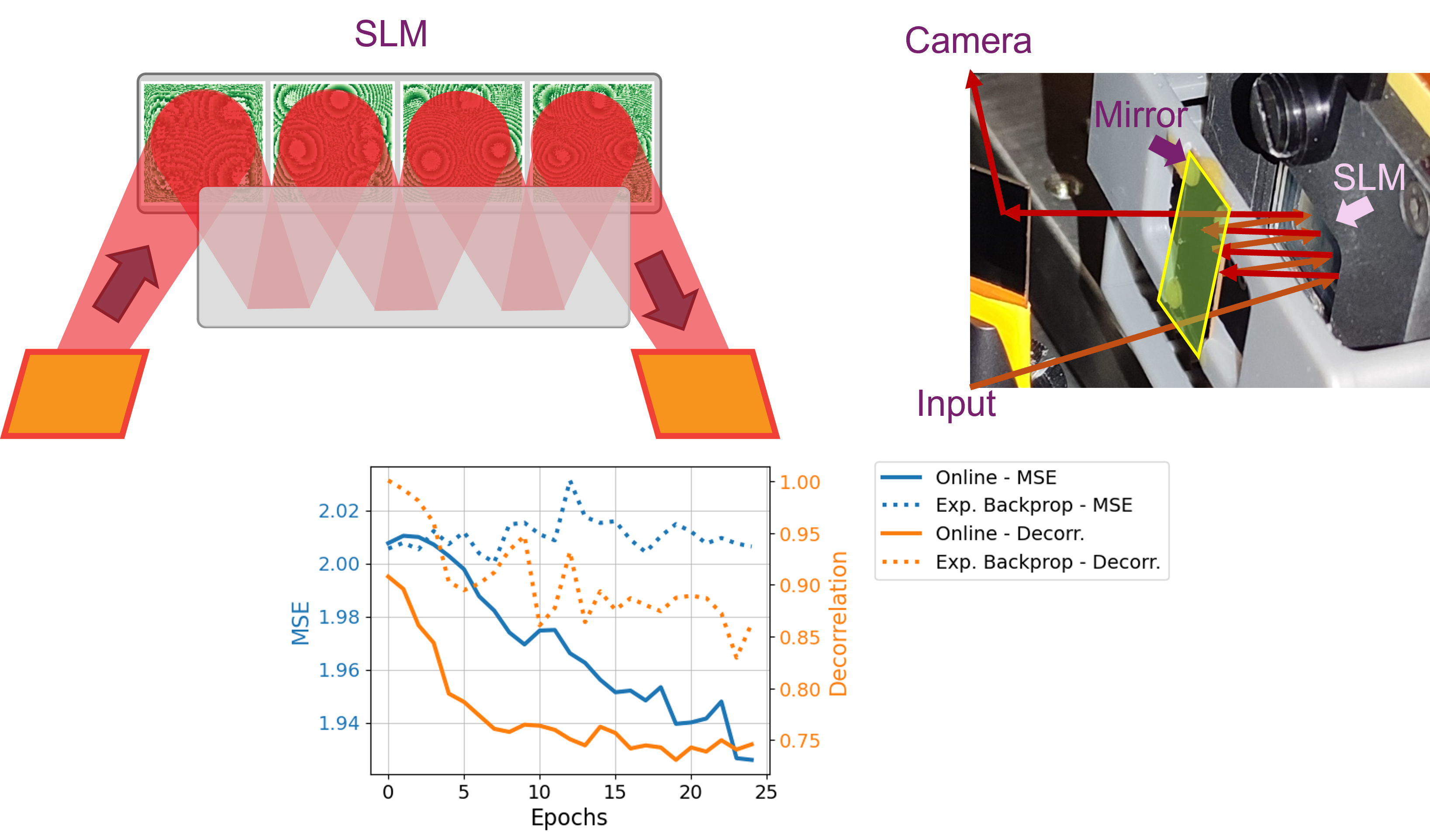}
  \medskip
  \caption{Schematic representation and photograph of the experimental system. Denoising error (MSE) and the decorrelation between the experimental system and the DT are plotted for online and only experimental backpropagation based trainings of the experimental system.}
  \label{fig:experimental_system_results}
  \medskip
\end{figure}

\subsection{Scalability and Efficiency Outlook on Optical Diffusion Models}\label{speed-energy}
In this section, we investigate the potential improvement in the proposed method's performance by using the same type of commercially available hardware and optimally scaling images onto the DMD and the camera. The widespread availability of newer optoelectronic technologies would enhance calculated accuracies further.

Considering fixed and passive modulation layers, the main energy and time consumption of the optical system stems from the input optical modulator, DMD and output array detector, CMOS camera. The DMD unit, Texas Instruments' DLP9500, can display 23,148 patterns per second at a resolution of $1920\times1080$, with an electrical consumption of 4.5 W at board level, including data transfer (i.e. looping back from the detector). When representing 8-bit images with the aggregation of 256 binary pixels as a superpixel, this device is capable of displaying 8-bit images with \textbf{\SI{24}{nJ/px}}. On the detection side, the CMOS camera (FLIR BFS-U3-04S2M-CS) can obtain 8-bit, 0.4 MP images at 522 fps rate with \SI{3}{W} power consumption, acquiring images at \textbf{\SI{14}{nJ/px}}. If we consider 1000 timesteps with $20\times20$ pixels images as in the rest of this study, the consumption due to optoelectric conversion and transfer would amount to \textbf{\SI{15}{mJ/image}}.

Another potential energy expense item could be the light intensity required to provide a sufficient signal-to-noise ratio (SNR) at the detector. The SNR, especially in the context of a digital device like a detector or an ADC (analog-to-digital converter), is typically calculated as $SNR(dB) = 6.02\times n+1.76$ where $n$ represents the number of bits. Considering the shot-noise-limited scenario, the required number of photons can be calculated using $N_{\text{input}} = \frac{SNR^2}{\eta_{\text{modulation layers}} \cdot \eta_{\text{detector}}}$ where $\eta_{\text{modulation layers}}$ accounts for a portion of light scattered out in passive layers and $\eta_{\text{detector}}$ accounts for the conversion efficiency of the detector. Assuming 90\% transmittance for each modulation layer and 50\% efficiency for the detector, and using an 850 nm wavelength to calculate the energy of each photon: $ E = \frac{hc}{\lambda} $, the required energy for light is only a few picojoules, which is negligible compared to the energy consumption of optoelectronic devices. 

With the optimal configuration of the ODU, where each pixel on the analytical model is precisely mapped one-to-one to the optoelectronic devices, the same set of hardware can generate images with a consumption of approximately \textbf{\SI{15}{mJ/image}} which is significantly lower than the 1.7 J required by conventional methods (see Appendix \ref{appendix:scaling-output}, making the ODU more than 100 times more energy-efficient.

\subsection{Training of Modulation Layers}\label{appendix:training}
As shown in Figure\ref{denoise-results}, we utilized 3 different datasets to demonstrate the proposed approach. 
\begin{itemize}
    \item First 3 digits from MNIST-digits
    \item First 3 classes from Fashion-MNIST
    \item 20000 "Clock" images from Quick, Draw! dataset
\end{itemize}

After being downsampled to $20\times20$, the images are used for 250 epochs to train the ODUs, using Adam optimizer with a learning rate of 0.006, which took $\sim10$ hours on an A100 GPU.

\subsection{Definition of Performance Metrics}\label{appendix:metric-definition}
\textbf{Mean Squared Error (MSE)}

\begin{equation}
\text{MSE} = \frac{1}{n} \sum_{i=1}^{n} (y_i - \hat{y}_i)^2
\end{equation}
where \( y_i \) are the true values and \( \hat{y}_i \) are the predicted values.

\textbf{Fréchet Inception Distance (FID)}

\begin{equation}
\text{FID}(x, g) = \| \mu_x - \mu_g \|^2 + \text{Tr}(\Sigma_x + \Sigma_g - 2 (\Sigma_x \Sigma_g)^{\frac{1}{2}})
\end{equation}
where \( (\mu_x, \Sigma_x) \) and \( (\mu_g, \Sigma_g) \) are the mean and covariance of the feature vectors of the real and generated data, respectively.

\textbf{Kernel Inception Distance (KID)}

\begin{equation}
\text{KID}(x, g) = \frac{1}{n(n-1)} \sum_{i \neq j} k(f(x_i), f(x_j)) + \frac{1}{m(m-1)} \sum_{i \neq j} k(f(g_i), f(g_j)) - \frac{2}{nm} \sum_{i, j} k(f(x_i), f(g_j))
\end{equation}
where \( k \) is a polynomial kernel and \( f \) is the Inception network function that extracts features.

\subsection*{Inception Score}

\begin{equation}
\text{IS}(G) = \exp \left( \mathbb{E}_{\mathbf{x} \sim p_g} \left[ D_{KL}(p(y|\mathbf{x}) \| p(y)) \right] \right)
\end{equation}
where \( p(y|\mathbf{x}) \) is the conditional label distribution given generated image \( \mathbf{x} \) and \( p(y) \) is the marginal distribution over all generated images.

\subsection{Code Availability}\label{code-availability}
The source code is available at \url{https://ioguz.github.io/opticaldiffusion/}.

\bibliography{OpticalDiffusion}

\end{document}